\documentclass[aps, prx, superscriptaddress, reprint, twocolumn, floatfix, preprintnumbers]{revtex4-2}
\usepackage{latexsym, amsmath, amsfonts, amssymb}
\usepackage{graphicx}
\usepackage{hyperref}
\hypersetup{colorlinks = true, linkcolor = blue, anchorcolor = blue, citecolor = blue, filecolor = blue, urlcolor = blue}
\usepackage{bbm}
\usepackage{mathrsfs}

\newcommand{\wt}{\widetilde}

\newcommand{\matht}[1]{{\ensuremath{\boldsymbol{#1}}}}
\newcommand{\tps}[2]{\texorpdfstring{#1}{#2}}
\newcommand{\novbox}[1]{\raisebox{0pt}[0pt][0pt]{#1}}

\newcommand{\eg}{\textit{e.g.}}

\newcommand{\ie}{\textit{i.e.}}

\newcommand{\etc}{\textit{etc}}

\newcommand{\nn}{\nonumber}

\newcommand{\be}{\begin{equation}} \newcommand{\ee}{\end{equation}}
\newcommand{\bea}{\begin{equation} \begin{aligned}} \newcommand{\eea}{\end{aligned} \end{equation}}

\newcommand{\cA}{\mathcal{A}}
\newcommand{\cB}{\mathcal{B}}
\newcommand{\cC}{\mathcal{C}}

\newcommand{\cF}{\mathcal{F}}

\newcommand{\cL}{\mathcal{L}}
\newcommand{\calL}{\mathscr{L}}
\newcommand{\cM}{\mathcal{M}}

\newcommand{\cP}{\mathcal{P}}

\newcommand{\cT}{\mathcal{T}}

\newcommand{\cW}{\mathcal{W}}

\newcommand{\cZ}{\mathcal{Z}}

\newcommand{\bQ}{\mathbb{Q}}
\newcommand{\bR}{\mathbb{R}}

\newcommand{\bZ}{\mathbb{Z}}

\newcommand{\fg}{\mathfrak{g}}

\newcommand{\fR}{\mathfrak{R}}

\DeclareMathOperator{\Tr}{Tr}

\newcommand{\bcenv}{\setlength{\leftmargini}{1.1em}\setlength{\abovedisplayskip}{4pt}\setlength{\belowdisplayskip}{4pt}}

\begin{document}

\preprint{SISSA 01/2024/FISI}

\title{Anomalies and gauging of \tps{\matht{U(1)}}{U(1)} symmetries}

\newcommand{\SISSA}{\affiliation{SISSA and INFN, via Bonomea 265, 34136 Trieste, Italy}}

\newcommand{\ICTP}{\affiliation{International Centre for Theoretical Physics (ICTP), Strada Costiera 11, 34151 Trieste, Italy}}

\author{Andrea Antinucci}
\SISSA

\author{Francesco Benini}
\SISSA
\ICTP

\begin{abstract}
We propose the Symmetry TFT for theories with a $U(1)$ symmetry in arbitrary dimension. The Symmetry TFT describes the structure of the symmetry, its anomalies, and the possible topological manipulations. It is constructed as a BF theory of gauge fields for groups $U(1)$ and $\bR$, and contains a continuum of topological operators. We also propose an operation that produces the Symmetry TFT for the theory obtained by dynamically gauging the $U(1)$ symmetry. We discuss many examples. As an interesting outcome, we obtain the Symmetry TFT for the non-invertible $\bQ/\bZ$ chiral symmetry in four dimensions.
\end{abstract}

\maketitle

\section{Introduction}

Symmetries and anomalies are some of the most basic, and yet robust and predictive, properties of a quantum field theory (QFT). Symmetries constrain spectra and correlation functions. Anomalies are exactly calculable and renormalization-group invariant, impose constraints on the low-energy dynamics (a famous example being 't~Hooft anomaly matching \cite{tHooft:1980}), and can explain a variety of phenomena. By inflow \cite{Wess:1971yu, Callan:1984sa}, the anomalies of a $d$-dimensional QFT can be described by a classical topological field theory (TQFT) of background fields in $d+1$ dimensions (this is also called an SPT phase, or an invertible TQFT \cite{Freed:2014iua}).

An improvement of this description, proposed in \cite{Gaiotto:2014kfa, Gaiotto:2020iye, Apruzzi:2021nmk, Freed:2022qnc}, captures at the same time the symmetries and anomalies of all possible variants (or global forms) of a seed theory, obtained by gauging finite subgroups of the symmetry group. This description goes under the name of Symmetry topological field theory (TFT). It is a nontrivial $(d+1)$-dimensional TQFT placed on a slab with two parallel boundaries. On one boundary it is coupled to the physical QFT$_d$. On the other boundary one prescribes a topological boundary condition. The claim is that there is a one-to-one correspondence between topological boundary conditions of the Symmetry TFT and global forms of the QFT. This proposal has been developed and verified in a variety of examples, that include the generalized symmetries of Ref.~\cite{Gaiotto:2014kfa}, and including cases with non-invertible (or categorical) symmetries \cite{Apruzzi:2022rei, Kaidi:2022cpf, Antinucci:2022vyk}. Moreover, the Symmetry TFT proved to be a useful tool to characterize both representations \cite{Bhardwaj:2023ayw} and anomalies, even of non-invertible symmetries \cite{Kaidi:2023maf, Zhang:2023wlu, Antinucci:2023ezl, Cordova:2023bja}, and has been applied to the classification of gapped phases \cite{Bhardwaj:2023fca}. So far it has been limited, though, to symmetries with a finite number of elements.

In this Letter we propose a construction of the Symmetry TFT for QFTs with continuous $U(1)$ symmetries. We provide a Lagrangian description of the Symmetry TFT, written in terms of gauge fields for group $\bR$ --- as opposed to $U(1)$. These theories are TQFTs with a continuum of operators, hence they constitute new mathematical objects with few properties that deviate from the standard well-known cases. Different boundary conditions of the TQFT correspond to topological manipulations of the boundary QFT, \ie, to gaugings of discrete symmetry subgroups, or alternatively of the whole $U(1)$ although in a ``flat'' way
\footnote{The flat gauging of a $U(1)$ $p$-form symmetry consists in summing over all flat bundles. It is a topological manipulation that does not introduce new degrees of freedom, and that yields a dual $(d{-}p{-}2)$-form symmetry $\bZ$.}.
This is enough to capture the anomalies of the theory, including the ordinary perturbative anomalies.

Physically, however, it is more interesting to understand the effect of the ordinary \emph{dynamical} gauging of a $U(1)$ symmetry, which involves the introduction of a new degree of freedom --- the photon --- coupled to the theory. In our setup, this is not a topological operation and therefore is not described by a boundary condition in  the original TQFT. Rather, it induces an operation that maps the Symmetry TFT to another one. We use this map to explicitly construct the Symmetry TFT for some interesting Abelian gauge theories in four dimensions. Interestingly, we observe that the various $(d+1)$-dimensional TQFTs can also be obtained from different boundary conditions in a $(d+2)$-dimensional TQFT that is a dynamical version of the anomaly polynomial \cite{Wess:1971yu}.

We provide a few selected examples of our construction. For instance, we give the Symmetry TFT description of the chiral anomaly in two and four dimensions, as well as the Symmetry TFT for a four-dimensional (4d) Abelian gauge theory with 2-group symmetry. We also provide a 3d example, illustrating interesting phenomena even in the absence of anomalies (that do not exist in odd dimensions). A particularly interesting outcome of our construction is the Symmetry TFT for the non-invertible $\bQ/\bZ$ symmetry that arises from a $U(1)$ chiral symmetry with ABJ (Adler--Bell--Jackiw) anomaly in 4d Abelian gauge theories \cite{Choi:2022jqy, *Cordova:2022ieu}. Our theory resembles, but is different from, the one discussed in \cite{Damia:2022bcd}. Constructions of $U(1)$ symmetry defects in string theory have been proposed in \cite{Cvetic:2023plv}.

In our notation, the Symmetry TFT lives in $d+1$ spacetime dimensions and has action $\cZ$, while the physical QFT is $d$-dimensional. We use Euclidean signature, so the path-integral weight is $e^{-\cZ}$. We use capital letters $A_p$ for $U(1)$ $p$-form gauge fields, and lowercase letters $a_p$ for $\bR$ $p$-form gauge fields. Sometimes we indicate a $p$-form symmetry with a superscript, \eg, $U(1)^{(p)}$.


\section{Symmetry TFT for \tps{\matht{U(1)}}{U(1)} symmetries}

The Symmetry TFT contains all the categorical data of the global symmetry of the boundary QFT$_d$. For the simple Abelian TQFTs considered in this paper, both the topological symmetry defects of the boundary QFT that (in the language of Ref.~\cite{Gaiotto:2014kfa}) generate the symmetry, and the charges that the operators can carry, are described by the bulk operators
\footnote{More generally, the Symmetry TFT contains complete information also on generalized charges, namely representations of $p$-form symmetries on $q$-dimensional objects, where $q\geq p$ \cite{Bhardwaj:2023ayw}.}.
A choice of boundary condition corresponds to a maximal set of mutually-transparent bulk operators (which we call a Lagrangian algebra $\calL$) that can terminate on the boundary. In other words, the boundary condition sets those operators to be trivial on the boundary. The endpoints (or more generally end-surfaces) of those operators correspond to the charged operators in the boundary theory, therefore $\calL$ is also the set of charges that the operators of the boundary theory can have.
On the contrary, we can produce topological operators of QFT$_d$ by laying the bulk operators on the topological boundary. Therefore the symmetry defects that generate the symmetry of the boundary theory are the operators of the Symmetry TFT modulo $\calL$.

We propose that the Symmetry TFT for continuous $U(1)$ symmetries, either 0-form or higher $p$-form, is a BF theory of gauge fields for gauge group $\bR$, as opposed to $U(1)$. To explain this point, we will first review the ordinary BF theory description of $\bZ_N$ gauge theory.

\medskip\noindent
\textbf{\matht{U(1)} gauge fields.} The $\bZ_N$ gauge theory in $d+1$ dimensions can be formulated as a BF theory of standard $U(1)$ gauge fields using the action \cite{Maldacena:2001ss, Banks:2010zn, Kapustin:2014gua}
\be
\label{U(1)/U(1) BF action}
\cZ = \frac{iN}{2\pi} \int_{X_{d+1}} B_{d-p} \wedge d A_p \,.
\ee
The gauge fields are not globally defined forms: they are patched using $U(1)$ gauge transformations $\delta A_p = d\lambda_{p-1}$ and $\delta B_{d-p} = d\lambda_{d-p-1}$. This leads to the Dirac quantization condition
\be
\frac1{2\pi} \int_{\gamma_{p+1}} dA_p \in \bZ \,,\qquad \frac1{2\pi} \int_{\gamma_{d-p+1}} dB_{d-p} \in \bZ \,,
\ee
where $\gamma_j$ are closed $j$-cycles. We can define extended operators
\be
\label{extended operators}
U_\alpha[\gamma_{d-p}] = e^{i\alpha \int_{\gamma_{d-p}} B_{d-p}} \,,\qquad W_\beta[\gamma_p] = e^{i\beta \int_{\gamma_p} A_p} \,,
\ee
and invariance under large gauge transformations requires $\alpha,\beta \in \bZ$.

In order to obtain the EOMs, it is convenient to decompose each gauge field into a representative of nontrivial $U(1)$ bundles, over which one has to sum, and a globally defined form describing fluctuations within a given bundle. Variations with respect to the fluctuations give $dA_p = 0$ and $dB_{d-p} = 0$, which guarantee that the operators (\ref{extended operators}) are topological. The sum over bundles produces delta functions imposing
\be
\frac{N}{2\pi} \int_{\gamma_p} \! A_p \in \bZ \,,\qquad \frac{N}{2\pi} \int_{\gamma_{d-p}} \! B_{d-p} \in \bZ \,.
\ee
This implies that the operators (\ref{extended operators}) with $\alpha \to \alpha+N$ or $\beta \to \beta+N$ are equivalent. Hence the theory has the following nonequivalent topological operators
\footnote{The theory might contain other topological operators, for instance condensates \cite{Roumpedakis:2022aik} or theta-defects \cite{Bhardwaj:2022lsg, *Bhardwaj:2022kot}.}:
\be
\label{operators in ZN BF}
U_n[\gamma_{d-p}] \,,\qquad W_m[\gamma_p] \,,\qquad \text{with } n,m \in \bZ_N \,.
\ee
The braiding between these operators can be computed by inserting one operator in (\ref{U(1)/U(1) BF action}) as a source, and then evaluating the VEV of the other one with the EOMs. The result for the braiding is
\be
\label{braiding in BF}
\cB(\alpha,\beta) = \exp\biggl( \frac{2\pi i}N \, \alpha \beta \biggr) \;.
\ee
This is the phase picked up by correlation functions when an operator $U_\alpha$ is moved across an operator $W_\beta$. The data (\ref{operators in ZN BF}) and (\ref{braiding in BF}) characterizes the $\bZ_N$ gauge theory.

There are various topological boundary conditions.
{\bcenv\begin{itemize}

\item A natural boundary condition is that the operators $W_m$ (with $m \in \bZ_N$) can terminate on the boundary. The defects $U_n$ with $n\in\bZ_N$ can lie on the boundary and play the role of symmetry defects for a
\be
\text{$\bZ_N$ $(p \,{-}1)$-form symmetry} \,.
\tag{a}
\ee
We can express the condition as a Dirichlet boundary condition for $A_p$ which then plays the role of the background field for the $(p-1)$-form symmetry. Anomalies are obtained by adding topological terms to $\cZ$ written in terms of $A_p$. These terms affect the EOMs and in general change the list of topological operators, the braiding, the possible boundary conditions, \etc.

\item Another boundary condition is to let the defects $U_n$ (with $n \in \bZ_N$) terminate on the boundary. This gives
\be
\text{$\bZ_N$ $(d{-}p{-}1)$-form symmetry}
\tag{b}
\ee
on the boundary. Indeed, the two symmetries are related by the discrete gauging of $\bZ_N$ in the boundary theory.
\end{itemize}}
\noindent
Other boundary conditions might exist, for instance when $N$ has divisors, or when $p = d-p$.

\medskip\noindent
\textbf{\matht{\bR} gauge fields.} In Ref.~\cite{Benini:2022hzx} it was found that the theory of a 2d free compact scalar --- that has $U(1)^2$ global symmetry --- admits a dual holographic description in terms of a certain 3d Chern--Simons theory of two gauge fields for group $\bR$, as opposed to $U(1)$. This suggests to study the Abelian BF theory with action
\be
\label{R/R BF action}
\cZ = \frac{i}{2\pi} \int_{X_{d+1}} b_{d-p} \wedge d a_p \,,
\ee
where $a_p$ and $b_{d-p}$ are gauge fields for the group $\bR$. This means that they are globally-defined forms, subject to small but not large gauge transformations. $\bR$-bundles are necessarily trivial, namely the Dirac quantization conditions collapse to
\be
\int_{\gamma_{p+1}} da_p = \int_{\gamma_{d-p+1}} db_{d-p} = 0 \,,
\ee
which is Stokes' theorem. We can always rescale $a_p$ or $b_{d-p}$ by a real constant, so that the overall coefficient in $\cZ$ is unphysical and we have fixed it to $1$.

The EOMs simply set $da_p=0$ and $db_{d-p}=0$, so that the operators $U_\alpha$ and $W_\beta$ are topological. Since there are no large gauge transformations, those operators are gauge invariant with no restrictions on $\alpha$ and $\beta$. Since there are no nontrivial bundles to sum over, there are no restrictions on the holonomies. We conclude that the theory has the following topological operators:
\be
\label{operators in R BF}
U_\alpha[\gamma_{d-p}] \,,\qquad W_\beta[\gamma_p] \,,\qquad \text{with } \alpha, \beta \in \bR \,.
\ee
The braiding is as in (\ref{braiding in BF}) but with $N=1$. Various topological boundary conditions are possible.
{\bcenv\begin{itemize}

\item Let all defects $W_\beta[\gamma_p]$ terminate on the boundary. They represent charged operators along $\partial\gamma_p$ with generic charges $\beta \in \bR$. All defects $U_\alpha[\gamma_{d-p}]$ can lie on the boundary and play the role of the symmetry defects for an
\be
\label{R (p-1)-form symmetry}
\text{$\bR$ $(p\,{-}1)$-form symmetry} \,.
\tag{c}
\ee
Such a symmetry indeed allows for generic charges. An example of a theory with this symmetry is a free $\bR$ $(p-1)$-form gauge field (for $p=1$ this is a free noncompact scalar).

\item Similarly, let all defects $U_\alpha[\gamma_{d-p}]$ terminate on the boundary. This describes an
\be
\label{R (d-p-1)-form symmetry}
\text{$\bR$ $(d{-}p{-}1)$-form symmetry}
\tag{d}
\ee
on the boundary. It is obtained from (\ref{R (p-1)-form symmetry}) by gauging the whole $\bR$ on the boundary (the Pontryagin dual to $\bR$ is $\bR$). This gauging, in order to be topological, needs to be \emph{flat}. This means that the field strength of the boundary gauge field is identically zero and we only sum over flat connections.

\item Let the defects $U_n$ and $W_m$ with $n,m \in \bZ$ terminate on the boundary. From (\ref{braiding in BF}), this choice constitutes a maximal set of mutually transparent operators and is thus a Lagrangian algebra. The coset classes of operators that can lie on the boundary are given by $U_\alpha$ and $W_\beta$ with $\alpha, \beta \in \bR/\bZ = U(1)$. Thus there are two factors, such that charged operators have integer charges while defects are valued in $U(1)$. This describes a
\be \label{Maxwell}
\text{$U(1)^{(p-1)} \times U(1)^{(d-p-1)}$ symmetry} \,.
\tag{e}
\ee
The two factors have a mixed anomaly. This is obtained from (\ref{R (p-1)-form symmetry}) by gauging a $\bZ$ subgroup of $\bR$ ($U(1)$ is the Pontryagin dual to $\bZ$) and the mixed anomaly follows from the fact that the exact sequence $0 \to \bZ \to \bR \to U(1) \to 0$ does not split \cite{Tachikawa:2017gyf}. An example of a theory with this symmetry is a free $U(1)$ $(p-1)$-form gauge field (for $p=1$ this is a free compact scalar, for $p=2$ a free photon), dual to a free $U(1)$ $(d-p-1)$-form gauge field. We refer to these theories as generalized Maxwell theories.

\end{itemize}}
\noindent
In the last case, one could more generally consider the defects $U_{nR}$ and $W_{mR^{-1}}$ for any real constant $R$. This would amount to rescaling the radii of the two $U(1)$ factors. The Symmetry TFT does not determine the actual value of the radius, but can compare the radii arising in two different topological boundaries. For instance, for $p=1$ we get a compact boson whose radius is $R$ times larger than the one in case \eqref{Maxwell}. Similarly, for $p=2$ the choice of $R$ corresponds to a rescaling of the electric charge in Maxwell's theory. The special case of a rescaling by $R = N \in \bZ$ also corresponds to gauging a $\bZ_N$ subgroup of $U(1)^{(d-p-1)}$. The cases (\ref{R (p-1)-form symmetry}) and (\ref{R (d-p-1)-form symmetry}) with symmetry $\bR$ correspond to the decompactification limits of the original field or of its dual, respectively.

\medskip\noindent
\textbf{\matht{U(1)/\bR} gauge fields.} Lastly, consider the case
\be
\label{U(1)/R BF action}
\cZ = \frac{i}{2\pi} \int_{X_{d+1}} \! b_{d-p} \wedge d A_p
\ee
where $A_p$ and $b_{d-p}$ are $U(1)$ and $\bR$ gauge fields, respectively. As before, the overall coefficient of $\cZ$ can always be set to $1$. This time Dirac's quantization conditions read $\frac1{2\pi} \!\int\! dA_p \in \bZ$ and $\int\! db_{d-p} = 0$. By the same arguments as before, one concludes that the theory has the following nonequivalent topological operators:
\be
U_\alpha[\gamma_{d-p}] \text{ with } \alpha \in \bR/\bZ = U(1) \,,\quad W_m[\gamma_p] \text{ with } m \in \bZ \,.
\ee
Some interesting boundary conditions are the following.
{\bcenv\begin{itemize}

\item Let the defects $W_m$ (with $m \in \bZ$) terminate on the boundary. Then the defects $U_\alpha$ with $\alpha \in U(1)$ can lie on the boundary and represent a
\be
\label{U(1) (p-1)-form symmetry}
\text{$U(1)$ $(p\,{-}1)$-form symmetry} \,.
\tag{f}
\ee

\item Let all defects $U_\alpha$ with $\alpha \in U(1)$ terminate on the boundary. Then the defects $W_m$ with $m \in \bZ$ can lie on the boundary and represent a
\be
\label{Z (d-p-1)-form symmetry}
\text{$\bZ$ $(d{-}p{-}1)$-form symmetry} \,.
\tag{g}
\ee
This is obtained from (\ref{U(1) (p-1)-form symmetry}) by the flat gauging of the whole $U(1)$. For \mbox{$p = d-1$}, an example of a theory with 0-form symmetry $\bZ$ is a scalar field $\phi$ with periodic potential, as in band theory.

\item Let the defects $W_{bm}$ with $m \in \bZ$ and $b>1$ an integer constant, as well as $U_{n/b}$ with $n \in \bZ_b$, terminate on the boundary. Then the nonequivalent defects that can lie on the boundary are $U_\alpha$ with $\alpha \in \bR / \bigl( \frac1b \bZ \bigr) \cong U(1)$ and $W_k$ with $k \in \bZ_b$. They represent a
\be
\text{$U(1)^{(p-1)} \times \bZ_b^{(d-p-1)}$ symmetry} \,.
\tag{h}
\ee
This is obtained from (\ref{U(1) (p-1)-form symmetry}) by gauging a $\bZ_b$ subgroup of $U(1)$. Indeed the charged operators have integer charges that are multiples of $b$. There is a mixed anomaly between $U(1)$ and $\bZ_b$ that follows from the short exact sequence $0 \to \bZ_b \to U(1) \,\to U(1) \to 0$.

\end{itemize}}

As a check, consider the case \eqref{Maxwell} that we derived from the Symmetry TFT \eqref{R/R BF action} using only $\bR$ gauge fields. It should also arise from the Symmetry TFT of two $U(1)$ symmetries with a mixed anomaly:
\be \label{eq:Alternative maxwell}
\cZ = \frac{i}{2\pi} \!\int\! \Bigl[ b_{d-p} \wedge dB_p + a_p \wedge dA_{d-p} - B_p \wedge dA_{d-p} \Bigr] .
\ee
Indeed the field $A_{d-p}$ can be integrated out producing a delta function that enforces $B_p = a_p$ (restricting $B_p$ to be an $\bR$ gauge field), and integrating out $B_p$ one reproduces the action (\ref{R/R BF action}). Alternatively, and more precisely, one can list the topological operators and compute their correlations functions. One realizes that $e^{in \!\int\! B_p}$ has identical correlation functions to those of $e^{i\alpha \!\int\! a_p}$ for $\alpha=n$, hence the two operators are identified. Similarly for $e^{im \!\int\! A_{d-p}}$ and $e^{i\beta \!\int\! b_{d-p}}$ for $\beta=m$. The alternative presentation \eqref{eq:Alternative maxwell} of the Symmetry TFT for a generalized Maxwell field will be important in Section~\ref{sec:dynamical gauging}.

\section{Examples}

Let us present a few interesting examples. Other examples, which require us to understand how to dynamically gauge a $U(1)$ symmetry from the point of view of the Symmetry TFT, are described in Section~\ref{sec:more examples}.

\subsection{Chiral anomaly in 2d}\label{sec:chiral 2d}

The Symmetry TFT for a two-dimensional theory with $U(1)$ 0-form symmetry and an 't~Hooft anomaly is obtained from the action (\ref{U(1)/R BF action}) with $d=2$ and $p=1$ by adding a term that describes the chiral anomaly:
\be
\label{Sym TFT for 2d chiral anomaly}
\cZ = \int_{X_3} \biggl[ \frac{i}{2\pi} \, b_1 \wedge dA_1 + \frac{ik}{4\pi} \, A_1 \wedge dA_1 \biggr] \,,
\ee
where $k$ is constrained to be integer (when $k$ is odd the theory requires a spin structure \cite{Dijkgraaf:1989pz}).

The theory has topological line operators given by
\be
U_{(n,\alpha)}[\gamma_1] = e^{i \!\int_{\gamma_1} \! ( n A_1 + \alpha \, b_1)}
\ee
with $n\in \bZ$ and $\alpha \in \bR$. The braiding between them is
\be
\label{3d braiding}
\cB\bigl[ (n_1, \alpha_1) , (n_2, \alpha_2) \bigr] = \exp\Bigl[ 2\pi i \bigl( n_1 \alpha_2 + n_2 \alpha_1 - k \alpha_1 \alpha_2 \bigr) \Bigr]
\ee
and the exponentiated spin is given by a quadratic refinement thereof:
\be
\theta_{(n,\alpha)} = \exp\Bigl[ 2\pi i \, \alpha \bigl( n - \tfrac k2 \, \alpha \bigr) \Bigr] \,.
\ee
The line $(k,1)$ has spin $\theta = (-1)^k$ and is a transparent fermion for $k$ odd. Both spin and braiding are invariant under the following identifications:
\be
\begin{cases} (n,\alpha) \,\sim\, (n + k ,\, \alpha + 1) &\text{for $k$ even,} \\ (n, \alpha) \,\sim\, (n+2k ,\, \alpha + 2) &\text{for $k$ odd.} \end{cases}
\ee

Let us discuss boundary conditions. First, we can let all lines $(n,0)$ terminate on the boundary. For $k$ even, this is a maximal set $\calL$ of mutually-transparent lines and they are all bosonic. For $k$ odd, we should also let the lines $(n,1)$ terminate on the boundary in order to have a maximal set $\calL$, and the extra lines have spin $-1$. Thus, the Lagrangian algebra $\calL$ is bosonic for $k$ even, and spin for $k$ odd. The nonequivalent topological line operators that can lie on the boundary are labeled by $\alpha \in \bR/\bZ \cong U(1)$, so this describes a $U(1)$ 0-form symmetry.

Let us use the lines $(0, \alpha)$ as representatives of the symmetry defect operators of the boundary theory. For $k$ even, the bosonic lines $(n,0)$ that end on the 2d boundary represent boundary local operators, and their charge measured by the braiding (\ref{3d braiding}) is $n$. For $k$ odd, the lines $(n,0)$ and $(n+k,1)$ that end on the boundary represent local operators with charge $n$ and spin $\pm1$, respectively. Thus the local operators can have arbitrary and independent integer charges and statistics.

It is possible to specialize to theories with a spin-charge relation, in which even-charge operators are bosonic and odd-charge operators are fermionic (\eg, a free complex Weyl fermion). In this case the Symmetry TFT (\ref{Sym TFT for 2d chiral anomaly}) is written in terms of a spin$_c$ connection $A_1$ \cite{Seiberg:2016rsg, Seiberg:2016gmd} whose Dirac quantization condition on 2-cycles is modified according to the second Stiefel--Whitney class of the manifold: $\frac1{2\pi} dA_1 = \frac12 w_2 \text{ mod }1$ (one also needs to add a gravitational term to the action). Since $A_1$ is not an ordinary connection but $2A_1$ is, gauge invariance restricts the operators $U_{(m,\alpha)}$ to have even $m$. This in turn implies the spin-charge relation for the boundary local operators: the endpoints of $(n,0)$ are bosonic for $n$ even, and the endpoints of $(n+k,1)$ are fermionic for $n$ odd.

It turns out that one can always formulate the Symmetry TFT as a bosonic theory, possibly with spin boundary conditions. Indeed the TQFT in (\ref{Sym TFT for 2d chiral anomaly}) with $k = k_\text{o}$ odd can be rewritten as the bosonic theory with $k = 4 k_\text{o}$, but with the fermionic $\bZ_2$ line $\cF = \bigl( 2k_\text{o}, \frac12 \bigr)$ gauged. One can implement such a gauging by including the fermionic $\bZ_2$ line $\cF$ in the spin Lagrangian algebra $\calL_\text{s} = \bigl\{ \bigl( 2n, \frac m2 \bigr) \bigr\}{}_{n,m \in \bZ}$ that describes the topological boundary condition, thus reproducing the original case. The line $\cF$ lied on the boundary generates the $\bZ_2^F$ fermion number symmetry $(-1)^F$. However there also exists a bosonic Lagrangian algebra $\calL_\text{b} = \bigl\{ (n, 0) \bigr\}$. It describes the bosonization of the original theory, obtained by gauging $(-1)^F$ on the boundary. The symmetry is still $U(1)$, but now it is an extension of the original $U(1)$ by the dual to $(-1)^F$. Thus the extended Symmetry TFT describes $U(1)^{(0)} \times \bZ_2^F$ in the original theory and their gaugings, including bosonization/fermionization
\footnote{We thank an anonymous referee for suggesting this.}.

When $k \neq 0$, the operators $U_{(0, \alpha)}$ are no longer mutually transparent and therefore there is no boundary condition corresponding to (\ref{Z (d-p-1)-form symmetry}). This is a manifestation of the chiral anomaly: the $U(1)^{(0)}$ symmetry of the boundary theory cannot be gauged. Indeed the anomaly of a continuous group is uniquely determined by the anomaly of all its discrete subgroups \cite{Thorngren:2020yht}. Therefore, a $U(1)$ symmetry has a perturbative anomaly if and only if there is an obstruction to its flat gauging. 

However, it might still be possible to gauge a discrete subgroup of $U(1)$. Consider the bosonic case of $k$ even. Given an integer $d$ that divides $k/2$, consider the set $\calL$ of line operators $\bigl( \frac{k}{2d} m + d \, \ell, \frac md \bigr)$ with $\ell,m \in \bZ$ (modulo identifications). Such operators have spin $\theta=1$ and are thus bosons. From (\ref{3d braiding}), a line that has trivial braiding with all elements of $\calL$ must be in $\calL$, thus $\calL$ is maximal and is a Lagrangian algebra. This shows that when $k/2$ is divisible by $d$, the $\bZ_d$ subgroup of $U(1)$ is anomaly free and can be gauged in the boundary theory. The nonequivalent topological lines that can lie on the boundary are $U_{(n,\alpha)}$ with $n \in \bZ_d$ and $\alpha \in \bR/\bigl( \frac1d \bZ \bigr)$. They represent a symmetry $U(1)\raisebox{0pt}[0pt]{$^{(0)}$} \times \bZ_d\raisebox{0pt}[0pt]{$^{(0)}$}$ with an 't~Hooft anomaly for $U(1)$ and a mixed anomaly between $U(1)$ and $\bZ_d$.

As a check, we can simply restrict to $\bZ_d$ the anomaly-inflow action $\cZ_\text{inflow} = \frac{ik}{4\pi} \!\int\! A \wedge dA$, where $A$ is seen as an extension from 2d to 3d of the background gauge field for the $U(1)$ symmetry \cite{Cordova:2018acb}. This is achieved by replacing
\be
\label{eq:mysterious algebra}
A \,\mapsto\, \frac{2\pi}d \, \cA \,,\qquad\quad \frac{dA}{2\pi} \,\mapsto\, \beta(\cA) \,,
\ee
where $\cA \in H^1(X; \bZ_d)$ is (an extension of) the background field for $\bZ_d$, while $\beta: H^1(X; \bZ_d) \to H^2(X; \bZ)$ is the Bockstein homomorphism associated to the exact sequence $0 \to \bZ \, \raisebox{0.45em}[0pt]{\makebox[0pt][l]{\scriptsize\hspace{0.8em}$d$}} \to \bZ \to \bZ_d \to 0$.
The inflow action reduces to $2\pi i \frac{k}{2d} \!\int\! \cA \cup \beta(\cA)$, where the integral is an integer modulo $d$, which is indeed an integer multiple of $2\pi$ whenever $d$ divides $k/2$ and thus $\bZ_d$ is anomaly free. The converse, to determine which $\bZ_d$ subgroups are actually anomalous, is a delicate issue \cite{Hsieh:2018ifc, Davighi:2020uab}: one should determine whether the reduced anomaly-inflow action is trivial when evaluated on the generator(s) of the relevant bordism group $\Omega^{SO}_3(B\bZ_d)$.

In the fermionic case of $k$ odd one can perform a similar analysis. Given $d$ that divides $k$ (in particular $d$ is odd), the set $\calL$ of line operators $\bigl( \frac kd \frac{d+1}2 m + d \, \ell, \frac md \bigr)$ labeled by $\ell, m \in \bZ$ (here $\frac{d+1}2 = 2^{-1}$ mod $d$) is a maximal set of mutually transparent lines with spins $\theta = (-1)^m = \pm1$, suggesting that the $\bZ_d$ subgroup can be gauged. This should be compared with the evaluation of the reduced inflow action on the generator(s) of the bordism group $\Omega^\text{spin}_3(B\bZ_d)$.

We ask whether there can be other more exotic topological boundary conditions. To be concrete, take $k=1$ and consider the following set:
\be
\cC = \bigl\{ \bigl(n, n + \ell \sqrt{2} \bigr) \bigm| n, \ell \in \bZ \bigr\} \,.
\ee
The set is closed under sum. Moreover, mapping the second component to the interval $[0,2)$ (by shifting also the first component), we obtain a dense irrational subset. The operators in $\cC$ have $\theta = (-1)^n$ and are all mutually transparent, however, the set is not maximal (hence it does not describe a topological boundary condition, but rather a topological interface) and in fact it cannot be made into a maximal set. A line mutually transparent with $\cC$ is of the form \novbox{$\bigl( m, m + \frac{h}{\sqrt2} \bigr)$} for some $m,h \in \bZ$. For $h$ odd, these lines are not contained in $\cC$, however they cannot be included in $\cC$ because they have nontrivial spin $\theta = i \, (-1)^{m+1}$. We have thus found an example of an Abelian algebra that cannot be completed into a Lagrangian algebra. Interestingly, this is a peculiarity of TQFTs with a continuum of lines, as can be intuitively understood from the necessity of taking the square root. Indeed a finite semi-simple modular tensor category admitting Lagrangian algebras is a Drinfeld center, and the condensation of an algebra of a Drinfeld center yields another Drinfeld center \cite{Davydov:2010kfz}. As we showed, this result is no longer true in the presence of a continuum of lines.

\subsection{\tps{\matht{U(1)}}{U(1)} symmetry in 3d}

In odd dimensions there are no anomalies for $U(1)$ symmetries. Nevertheless there are useful pieces of information encoded in the symmetry TFT. Here we consider the 3d case, hence the basic 4d Symmetry TFT describing a $U(1)$ symmetry is (\ref{U(1)/R BF action}) with $d=3$, $p=1$.
There are two more terms that can be added: a theta-term $\frac{i\theta}{8\pi^2} \!\int_{X_4} \! dA_1\wedge dA_1$ and a phi-term $\frac{i\phi}{8\pi^2} \!\int_{X_4} \! b_2 \wedge b_2$. In the absence of the latter, the theta-term is unphysical since it can be reabsorbed by a shift of $b_2$. Here we study the effect of the phi-term, while we leave the analysis of the most general action with both terms for the future.

The 4d TQFT we consider is 
\be
\label{4d Sym TFT action with phi}
\cZ = \frac{i}{2\pi} \int_{X_4} b_2 \wedge dA_1 + \frac{\phi}{i8\pi^2} \int_{X_4} b_2 \wedge b_2 \,.
\ee
This can be thought of as the noncompact version of the theory studied in Ref.~\cite{Kapustin:2014gua}. The gauge transformations are
\be
\label{gauge transformations 3d/4d case}
\delta b_2 = d\lambda_1 \,,\qquad \delta A_1 = d\rho_0 - \tfrac{\phi}{2\pi} \, \lambda_1 \,.
\ee
The gauge-invariant operators are the surfaces
\be
U_\alpha [\gamma_2] = e^{i\alpha \!\int_{\gamma_2} b_2}
\ee
and the (generically) non-genuine lines
\be
\label{eq:non genuine line 4d}
W_n[\gamma_1,D_2] = \exp \biggl( in \int_{\gamma_1} A_1 + i \, \frac{n\phi}{2\pi} \int_{D_2} b_2 \biggr) ,
\ee
both topological. Here $n\in\bZ$ while $D_2$ is a disk with $\partial D_2=\gamma_1$. A nontrivial correlator on the sphere is the braiding of $U_\alpha[\gamma_2]$ and $W_n[\gamma_1,D_2]$ in a configuration in which $\gamma_1$ and $\gamma_2$ link, hence $\gamma_2$ intersects $D_2$ at a point:
\be
\bigl\langle U_\alpha [\gamma_2] \, W_n[\gamma_1,D_2] \bigr\rangle = e^{2\pi i \alpha n} \,.
\ee
We read off that $\alpha \sim \alpha+1$. This implies that if $\frac{n\phi}{2\pi}$ is an integer in \eqref{eq:non genuine line 4d}, there is no dependence of $W_n$ on the disk $D_2$: choosing a different disk would give an operator which differs by $U_\alpha[\gamma_2]$ for some integer $\alpha$ and where $\gamma_2$ is the union (with opposite orientations) of the two disks, hence that difference is trivial. In particular we read off that $\phi$ is a periodic parameter:
\be
\label{periodicity of phi}
\phi \sim \phi + 2\pi \,.
\ee
An interesting observation is that, while for irrational values of $\frac{\phi}{2\pi}$ the theory has no genuine lines, hence it is essentially trivial in the bulk
\footnote{Here by trivial we mean that the theory is an invertible TQFT, also known as an SPT phase. Such a theory has endable surface operators (\eg, disks) whose intersections produce nontrivial phases. See also Ref.~\cite{Argurio:2024oym}},
when $\phi=2\pi p/q$ with $p,q \in \bZ$ and $\gcd(p,q)=1$, the lines $W_{mq}$ are genuine.

Let us study topological boundary conditions. First, we can let all surfaces $U_\alpha$ terminate on the boundary. This corresponds to a Dirichlet boundary condition for $b_2$. Since the gauge transformations of $b_2$ are forced to vanish at the boundary, there we can construct the genuine line operators $\cW_n = e^{in \!\int\! A_1}$ (equivalently, since $b_2$ is a background field at the boundary, the dependence of $W_n$ on it can be removed by a counterterm). This boundary condition thus describes a $\bZ$ 1-form symmetry, whose topological operators are the lines $\cW_n$ and whose charged operators are the endlines of $U_\alpha$ on the boundary. The phi-term represents an anomaly for this 1-form symmetry, and if $\frac{\phi}{2\pi}$ is irrational there are no non-anomalous subgroups of $\bZ$. Indeed in this case the bulk does not have other genuine topological operators besides $U_\alpha$, hence there are no other topological boundary conditions. In particular, there is no boundary with a $U(1)$ symmetry.

On the contrary, consider the case $\frac{\phi}{2\pi} = p/q \in \bQ$. The lines $W_{qm}$ with $m\in\bZ$ are genuine, as we noticed, and we can let them terminate on the boundary. In order to have a maximal set of mutually-transparent objects, we should let the surfaces $U_{\alpha}$ with $\alpha = l/q$ and $l \in \bZ$ terminate on the boundary as well
\footnote{In the presence of non-genuine surfaces, Ref.~\cite{Argurio:2024oym} discussed a generalized notion of Lagrangian algebra that must also include them. One can check that the boundary condition described here can be appropriately enriched with the inclusion of non-genuine surface operators.}.
This boundary condition describes the symmetry \novbox{$\bZ_q{}^{(1)} \times U(1)^{(0)}$} with a mixed anomaly and a pure anomaly for \novbox{$\bZ_q{}^{(1)}$}, and is obtained from the previous one by gauging the subgroup $q\bZ \subset \bZ$ of the 1-form symmetry, which is non-anomalous in this case. The symmetry $U(1)^{(0)}$ is generated by $U_\alpha$ with identification $\alpha \sim \alpha + 1/q$ (the reduced range is due to the boundary condition). The local operators $\cM_m$ charged under $U(1)^{(0)}$ are the endpoints of the genuine lines $W_{qm}$.

To get some intuition on the nature of the QFTs described by this boundary condition, let us present a Lagrangian example. On a manifold with boundary, the variation of the action (\ref{4d Sym TFT action with phi}) under a gauge transformation (\ref{gauge transformations 3d/4d case}) generates a boundary term:
\be
\label{gauge variation 4d Sym TFT}
\delta \cZ = \frac{i}{2\pi} \!\int_{\partial X_4} \biggl[ \lambda_1 \wedge dA_1 - \frac{\phi}{4\pi} \, \lambda_1 \wedge d\lambda_1 \biggr] \,,
\ee
where the boundary value of $A_1$ is fixed. This can be canceled by edge modes. Since the local operators $\cM_m$ charged under \novbox{$U(1)^{(0)}$} are the endpoints of $W_{qm}$, the corresponding background field is $qA_1$ at the boundary. Since the line operators charged under $\bZ_q{}^{(1)}$ are the endlines of $U_{l/q}$, the background field is $b_2/q$. At the boundary we can place the Chern--Simons theory $U(1)_{pq}$:
\be
S_\partial = -i \int_{\partial X_4} \biggl[ \frac{pq}{4\pi} \, \cB \wedge d\cB + \frac1{2\pi} (qA_1) \wedge d\cB \biggr] \,.
\ee
This theory has 1-form symmetry $\bZ_{pq} = \bZ_p \times \bZ_q$, but we only consider the $\bZ_q$ subgroup which has anomaly $p$ \cite{Hsin:2018vcg}. The generator of $\bZ_q$ acts as \novbox{$\delta \cB = \frac1q \lambda_1$} so that the variation of $S_\partial$ cancels (\ref{gauge variation 4d Sym TFT}). The Chern--Simons theory also has a magnetic $U(1)$ 0-form symmetry with current $* \, d\cB$, which is coupled to the background field $(qA_1)$. The local operators $\cM_m$ are monopoles. In this example the current is trivial at separated points, but it still has an interesting contact term \cite{Closset:2012vp} (to make the current nontrivial at separated points, one could consider Maxwell--Chern--Simons theory instead). Because of the 4d EOM, we can identify \novbox{$dA_1 = -p \, \bigl( \frac1q b_2 \bigr)$} with the background for the 1-form symmetry, hence the theory is coupled to the latter as well, in a subtle way.

Another possibility is to use the topological theory $\cA^{q,p}\bigl[ \frac1q dA_1 \bigr]$ \cite{Hsin:2018vcg} as boundary theory. Because of the identifications among those theories, this in particular shows that $p \sim p+q$ in accord with (\ref{periodicity of phi}).

\subsection{Chiral anomaly in 4d}
\label{sec:chiral 4d}

For a four-dimensional theory with a $U(1)$ 0-form symmetry and an 't~Hooft anomaly, the Symmetry TFT is
\be
\label{SymTFT for 4d chiral anomaly}
\cZ = \int_{X_5} \biggl[ \frac{i}{2\pi} \,  b_3 \wedge dA _1 + \frac{ik}{24\pi^2} \, A _1 \wedge dA_1 \wedge dA_1 \biggr]
\ee
where $A_1$ is a $U(1)$ gauge field while $b_3$ is an $\bR$ gauge field. The parameter $k$ is an integer in fermionic theories, while in general bosonic theories it should be a multiple of 6.

The theory has topological line and surface operators
\bea
W_n[\gamma_1] &= e^{in \!\int_{\gamma_1} \! A_1} \,, \\
U_{(\beta,m)}[\gamma_3] &= \exp \biggl[ i \!\int_{\gamma_3} \! \Bigl( \beta \, b_3 + \frac{m}{4\pi} \, A_1 \wedge dA_1 \Bigr) \biggr]
\eea
with $n,m \in \bZ$. The quantization of $m$ corresponds to spin-Chern--Simons theories. The perturbative EOMs $dA_1 = db_3 = 0$ guarantee that these operators are topological. The observables of the theory include the linking of $W_n$ with $U_{(\beta,m)}$, and the triple-linking between three operators $U_{(\beta_i,m_i)}$ on surfaces \novbox{$\gamma_3{}^{(i)}$}. The latter probes the linking between the intersection \novbox{$\widetilde{\gamma}_1 = \gamma_3{}^{(1)}\cap \gamma_3{}^{(2)}$} and $\gamma_3{}^{(3)}$ (one can show that this is symmetric in the three surfaces). The braiding can be defined as the following expectation value on the sphere:
\be
\label{eq:braiding 5d}
\bigl\langle W_n[\gamma_1] \, U_{(\beta,m)}[\gamma_3] \bigr\rangle = \exp \bigl( 2 \pi i \, n \beta \operatorname{Lk}(\gamma _1,\gamma_3) \bigr) \,,
\ee
where $\operatorname{Lk}$ is the geometric linking number. Alternatively, it is the phase picked up by generic correlation functions when $W_n$ is moved across $U_{(\beta,m)}$. The triple-linking number on the sphere is the following expectation value:
\begin{align}
\label{eq:triple linking}
& \bigl\langle U_{(\beta _1,m_1)} \bigl[ \gamma _3^{(1)} \bigr] \, U_{(\beta _2,m_2)} \bigl[ \gamma _3^{(2)} \bigr] \, U_{(\beta _3,m_3)} \bigl[ \gamma _3^{(3)} \bigr] \bigr\rangle \\[.2em]
&= \exp \bigl[ 2\pi i \bigl( m_1\beta _2\beta_3+m_2\beta _1\beta _3+m_3\beta _1\beta _2-k\beta _1\beta _2\beta _3 \bigr) \text{\L} \bigr] \nn
\end{align}
where \novbox{$\text{\L} = \operatorname{Lk}\bigl( \wt\gamma_1, \gamma_3{}^{(3)} \bigr)$}.
Note that, differently from the 2d case, there is no operator $U_{(\beta,m)}$ with trivial triple-linking with all other pairs of operators. Hence there are no identifications of labels in this case.

Using (\ref{eq:braiding 5d}) and (\ref{eq:triple linking}) we can look for topological boundary conditions corresponding to the condensation of a ``Lagrangian algebra''. By this we mean a set of line and surface operators which are: (i) closed under fusion, (ii) mutually transparent with respect to both braiding and triple linking, and (iii) maximal in the sense that any operator transparent with the set belongs itself to the set. These conditions guarantee that the Lagrangian algebra can be condensed, and the result is the trivial TQFT. We find the following possibilities.

First, the lines $W_n$ and the surfaces $U_{(\beta = j,m)}$ with $n, j, m \in \bZ$ are mutually transparent and maximal. Condensing all of them we obtain the boundary condition for a theory with $U(1)$ 0-form symmetry. Its nonequivalent topological symmetry operators are $U_{(\beta,0)}[\gamma _3]$ where $\beta \in \bR/\bZ$ with $\beta \sim \beta +1$ because of the condensation.

Second, if $k=0$, another Lagrangian algebra is given by all the operators $U_{(\beta,0)}$. This correspond to a 2-form symmetry $\bZ$ whose topological symmetry operators are the lines  $W_n$ with $n\in \bZ$, as in the general case (\ref{Z (d-p-1)-form symmetry})
\footnote{The boundary theory also inherits the topological surfaces $U_{(0,m)}[\gamma_3]$ which are theta-defects \cite{Bhardwaj:2022lsg, *Bhardwaj:2022kot} constructed out of the lines $W_n[\gamma_1]$.}.
Such a boundary condition is obtained from the previous one by flat gauging of the $U(1)$ symmetry on the boundary. If $k\neq 0$, however, this algebra does not exist, consistently with the statement that the Symmetry TFT describes an anomalous symmetry.

Lastly, given an integer $d$ such that $3d$ divides $k$, one can construct a Lagrangian algebra made of
\be
W_{d \rho} \,,\quad U_{(\beta,m)} \text{ with } (\beta,m) = \biggl( \frac\nu{d} ,\, \frac{k}{3d} \, \nu + d^2 \mu \biggr)
\ee
and labeled by $\rho,\nu, \mu \in \bZ$. The endpoints of the lines are the charged objects. Since the charges are multiples of $d$, we conclude that this boundary corresponds to gauging the non-anomalous $\bZ_d$ subgroup of $U(1)^{(0)}$. The symmetry on this boundary is $U(1)^{(0)} \times \bZ_{\novbox{\scriptsize$d$}}^{(2)}$ where the first factor is the $\bZ_d$ quotient of the original symmetry, while the second factor is the dual 2-form symmetry generated by the lines $W_p$ with $p \in \bZ_d$.

\section{Non-topological manipulations}
\label{sec:dynamical gauging}

The Symmetry TFT for $U(1)$ symmetries that we discussed is the straightforward generalization of the discrete case. Its topological boundaries correspond to \emph{topological} manipulations that use the $U(1)$ symmetry. These include gauging discrete subgroups, possibly with discrete torsion, as well as the flat gauging of the whole $U(1)$. However, differently from the discrete ones, continuous $U(1)$ symmetries also allow for non-topological, \emph{dynamical} manipulations --- such as coupling to a dynamical photon --- that introduce new degrees of freedom. For instance, a 4d free complex scalar field has a $U(1)^{(0)}$ symmetry and its Symmetry TFT is \eqref{U(1)/R BF action} with $p=1$, $d=4$. By gauging dynamically the $U(1)$ we obtain scalar QED which has a $U(1)^{(1)}$ symmetry, hence its Symmetry TFT is again \eqref{U(1)/R BF action} but with $p=2$, $d=4$.  We see that dynamical manipulations are not described by different topological boundary conditions of the same Symmetry TFT, but rather by a map between two different Symmetry TFTs
\footnote{Recently Ref.~\cite{Kobayashi:2021jsc} constructed 4d TQFTs coupled with a non-flat $U(1)$ \emph{background}. It would be interesting to explore the relation between making this background dynamical and our approach.}.
As we will argue, this is a controlled operation.

\subsection{Gauging dynamically a \tps{\matht{U(1)}}{U(1)}}

For concreteness we are going to focus on 0-form symmetries, but the generalization to higher forms is straightforward. The initial Symmetry TFT is \eqref{U(1)/R BF action} with $p=1$. The idea is to add to it the Symmetry TFT of a $d$-dimensional photon, and to couple the two TFTs in the bulk in a way that reproduces the coupling of the current to the gauge field on the boundary. It is convenient to use the alternative formulation \eqref{eq:Alternative maxwell} for the Symmetry TFT of the photon, which we report here for convenience:
\be
\cZ = \frac{i}{2\pi} \!\int \Bigl[ g_{d-2}\wedge dG_2 +f_2\wedge dF_{d-2}-G_2\wedge dF_{d-2} \Bigr] .
\ee
The coupling to the fields appearing in the ``matter'' part $\frac{i}{2\pi} \!\int b_{d-1}\wedge dA_1$ must be such that, on the boundary, the Wilson lines of the Maxwell field are endable on the local operators charged under $U(1)^{(0)}$. The Wilson lines are the endlines of the surfaces of $G_2$, while the above-mentioned local operators are the endpoints of the lines of $A_1$. Hence, the coupling must allow the surfaces of $G_2$ to end on the lines of $A_1$. We are led to the total action:
\begin{align}
\cZ &= \frac{i}{2\pi} \!\int_{X_{d+1}} \Bigl[ b_{d-1} \wedge dA_1 + b_{d-1} \wedge G_2 \\
&\qquad\quad + g_{d-2} \wedge dG_2 + f_2 \wedge dF_{d-2} - G_2 \wedge dF_{d-2} \Bigl]. \nn
\end{align}
Because of the added coupling $b_{d-1}\wedge G_2$, the standard gauge transformation of $G_2$ must also act on $A_1$:
\be
\delta G_2 = d\lambda_1 \,,\qquad \delta A_1 = - \lambda_1 \,.
\ee
Hence the lines of $A_1$ are no longer gauge invariant, but we can construct the non-genuine line operators
\be
L_n[\gamma_1, D_2] = e^{in \!\int_{\gamma_1} \! A_1 \,+\, in \!\int_{D_2} \! G_2}
\ee
where $\partial D_2 = \gamma_1$. This implies that all the surfaces of $G_2$ can end, and on the boundary all Wilson lines can be cut open, achieving the coupling of the photon to matter.

Similarly, the gauge transformations of $b_{d-1}$ must act on $g_{d-2}$:
\be
\delta b_{d-1} = d\eta_{d-2} \,,\qquad \delta g_{d-2} = (-1)^d \, \eta_{d-2} \,.
\ee
As a consequence the surfaces of $g_{d-2}$ are not gauge invariant, and must be attached to a disk where $b_{d-1}$ is integrated: 
\be
U_\alpha[\gamma_{d-2}, D_{d-1}] = e^{i\alpha \!\int_{\gamma_{d-2}} g_{d-2} \,-\, (-1)^d i \alpha \!\int_{D_{d-1}} \! b_{d-1} } .
\ee
This also has a clear physical interpretation. The operators $\exp \bigl( i\alpha \!\int b_{d-1} \bigr)$ that used to generate on the boundary the global $U(1)$ 0-form symmetry we are gauging, can now be opened and trivialized. Gauging a symmetry trivializes the topological operators that generate it.

By inspection we notice that the only genuine operators that cannot be opened are the surfaces of $f_2$ and $F_{d-2}$. All the other ones become trivial because of the coupling $b_{d-1}\wedge G_2$. Therefore the theory is equivalent to
\be
\label{eq:action dinamical gauging}
\cZ = \frac{i}{2\pi} \!\int_{X_{d+1}} \! f_2 \wedge dF_{d-2} \,.
\ee
This is nothing but the action \eqref{U(1)/R BF action} with $p=2$, and it describes the dual magnetic symmetry appearing after the dynamical gauging.

We also notice that going from $\frac{i}{2\pi} \!\int b_{d-1} \wedge dA_1$ to \eqref{eq:action dinamical gauging} can be understood as the replacement:
\be
\label{eq:procedure}
dA_1 \,\mapsto\, f_2 \,,\qquad\quad b_{d-1} \,\mapsto\, dF_{d-2} \,.
\ee
The first substitution means that $A_1$, which was previously flat, can now describe a curved background with field strength $f_2$, which is free --- hence dynamical --- at the boundary. The second substitution means that the previous (Hodge dual) current $b_{d-1}$ of the $U(1)^{(0)}$ symmetry is trivialized in terms of a form of lower degree.

The prescription \eqref{eq:procedure} is the basic ingredient to construct maps between different Symmetry TFTs, implementing the dynamical manipulations. In Section~\ref{sec:more examples} we will apply it to more complicated examples in order to construct the Symmetry TFTs of various symmetry structures.

\subsection{The Anomaly Polynomial TFT}

It is important to emphasize that, while for discrete symmetries the Symmetry TFT is unique and its topological boundaries describe all possible manipulations, here we need a step further. Indeed we find distinct Symmetry TFTs, related by the map \eqref{eq:procedure}, that describe the dynamical manipulations, while each Symmetry TFT admits various topological boundaries that describe the topological manipulations. We observe, however, that one can construct a (unique) $(d+2)$-dimensional TQFT whose $(d+1)$-dimensional topological boundaries correspond to the distinct Symmetry TFTs. We dub this the \emph{Anomaly Polynomial} TFT. It is written entirely in terms of $\bR$ gauge fields, which can be identified with the field strengths of the gauged symmetries in the various theories related by dynamical manipulations. 

We illustrate the idea in the simplest example of a 2d theory with a single $U(1)^{(0)}$ symmetry. Denoting by $k\in \bZ$ the anomaly, the Anomaly Polynomial TFT is a 4d TQFT with action:
\be
\label{eq:anomaly polynomial TFT 2d}
\cP = \frac{i}{2\pi} \!\int_{X_4} \biggl[ g_1 \wedge d f_2 + \frac{k}{2} \, f_2 \wedge f_2 \biggl]
\ee
and gauge transformations $\delta g_1 = d\rho_0 - k \lambda_1$, $\delta f_2 = d\lambda_1$. On a manifold with boundary the gauge variation produces a boundary term. It is not uniquely determined because we have the freedom to add a boundary term proportional to $g_1 \wedge f_2$, but independently of this choice the boundary gauge variation cannot be canceled unless we couple the 4d theory to a 3d theory of edge modes.

Consider first the case $k=0$. The gauge variation of \eqref{eq:anomaly polynomial TFT 2d} produces the boundary term
\be
\label{eq:boundary gauge variation 1}
\delta \cP = - \frac{i}{2\pi} \!\int_{\partial X_4} d\rho_0 \wedge f_2 \,.
\ee
In order to cancel it, we place on the boundary the following 3d TQFT of edge modes with a 1-form symmetry coupled to $f_2$, whose boundary value we regard as a background field:
\be
\label{eq:boundary1}
\cZ = \frac{i}{2\pi} \! \int_{\partial X_4} \Bigl[ b_1 \wedge dA_1 - b_1 \wedge f_2 \Bigr] \,.
\ee
The 1-form symmetry acts on the lines $e^{in \!\int\! A_1}$ and shifts $\delta A_1 = \lambda_1$. The gauge variation \eqref{eq:boundary gauge variation 1} is canceled by imposing the gluing condition $g_1 |_\partial = - b_1$, which is the boundary EOM from the variation of the total action with respect to $f_2$, and is compatible with the gauge transformation $\delta b_1 = - d\rho_0$. Turning off the background $f_2$, we recognize \eqref{eq:boundary1} as the Symmetry TFT for a 2d theory with $U(1)$ 0-form symmetry.

There is another boundary theory we can use. It is more cleanly presented if we first add to (\ref{eq:anomaly polynomial TFT 2d}) the boundary term $g_1\wedge f_2$, which is equivalent to recasting the Anomaly Polynomial TFT as $\cP' = \frac{i}{2\pi} \!\int_{X_4} \! dg_1 \wedge f_2$. The gauge variation produces the boundary term
\be
\label{eq:boundary gauge variation 2}
\delta \cP' = \frac{i}{2\pi} \!\int_{\partial X_4} g_1 \wedge d\lambda_1 \,.
\ee
Now we regard the boundary value of $g_1$ as a background field for the 0-form symmetry of a boundary 3d TQFT:
\be
\label{eq:boundary2}
\cZ' = \frac{i}{2\pi} \!\int_{\partial X_4} \Bigl[ h_2 \wedge d\Theta_0 - h_2 \wedge g_1 \Bigr] \,.
\ee
The 0-form symmetry shifts $\delta\Theta_0 = \rho_0$, and the variation \eqref{eq:boundary gauge variation 2} is canceled by imposing the gluing condition $f_2 |_\partial = h_2$, which follows for the variation of the total action with respect to $g_1$ and is compatible with the gauge transformation $\delta h_2 = d\lambda_1$. We recognize \eqref{eq:boundary2} as the Symmetry TFT for a $U(1)^{(-1)}$ symmetry \cite{Cordova:2019jnf}, related in 2d to the $U(1)^{(0)}$ symmetry by dynamical gauging. It corresponds to a shift of the theta angle \novbox{$\frac{\theta}{2\pi} \!\int_{X_2} \! \cF$} for the dynamical $U(1)$ gauge field. We observe indeed that turning off background fields, \eqref{eq:boundary2} is obtained from  \eqref{eq:boundary1} by using the map \eqref{eq:procedure} that implements dynamical gauging in the Symmetry TFT.

In the case with anomaly $k \neq 0$, a gauge variation produces the boundary term
\be
\delta \cP = \frac{i}{2\pi} \!\int_{\partial X_4} \biggl[ - \bigl( d\rho_0 - k \lambda_1 \bigr) \wedge f_2 + \frac{k}{2} \, \lambda_1 \wedge d\lambda_1 \biggr] .
\ee
It is canceled by the following modification of \eqref{eq:boundary1}:
\be
\label{eq:boundary1'}
\cZ = \frac{i}{2\pi} \!\int_\partial \biggl[ b_1 \wedge dA_1 + \frac{k}{2} A_1 \wedge dA_1 - \bigl( b_1 + k A_1 \bigr) \wedge f_2 \biggr].
\ee
The current in parenthesis that multiplies the background field $f_2$ is $\partial \cZ / \partial(dA_1)$. One uses the transformations $\delta A_1 = \lambda_1$, $\delta b_1 = -d\rho_0$ as well as the gluing condition $g_1 |_\partial = - b_1 - kA_1$ that follows from varying the total action with respect to $f_2$. With the background field $f_2$ off, we recognize \eqref{eq:boundary1'} as the Symmetry TFT (\ref{Sym TFT for 2d chiral anomaly}) for a $U(1)^{(0)}$ symmetry with chiral anomaly $k$ in 2d. It turns out that the other boundary theory \eqref{eq:boundary2} for $k=0$ cannot be modified in any way to cancel the gauge variation once $k\neq 0$, hence becoming an inconsistent boundary condition. This is a manifestation of the 't~Hooft anomaly, from the Anomaly Polynomial TFT viewpoint.

There is an analogous story for $U(1)$ symmetries in 4d, for which the Anomaly Polynomial TFT takes the form:
\be
\cP = \frac{i}{2\pi} \!\int_{X_6} \biggl[ g_3 \wedge df_2 - \frac{k}{12\pi} \, f_2 \wedge f_2 \wedge f_2 \biggr] \,,
\ee
with transformations $\delta g_3 = d\rho_2 + \frac{k}{2\pi} \lambda_1 \wedge f_2 + \frac{k}{4\pi} {\lambda_1 \wedge d\lambda_1}$ and $\delta f_2 = d\lambda_1$.
For $k=0$ there are two possible boundary theories, which are the Symmetry TFTs for a 0-form and 1-form $U(1)$ symmetries, respectively, related by dynamical gauging. For $k\neq 0$, instead, only the first one is consistent and it takes the form:
\begin{align}
\cZ &= \frac{i}{2\pi} \!\int_{\partial X_6} \biggl[ b_3 \wedge dA_1 + \frac{k}{12\pi} \, A_1 \wedge dA_1 \wedge dA_1 \\
&\qquad\qquad - \biggl( b_3 + \frac{k}{4\pi} \, A_1 \wedge dA_1 - \frac{k}{4\pi} A_1 \wedge f_2 \biggr) \wedge f_2 \biggr] \nn
\end{align}
where the terms in parenthesis describe the coupling of the 1-form symmetry to the background $f_2$ and correspond to $\partial\cZ/\partial(dA_1)$. One finds the gluing condition $g_3 |_\partial = - b_3 - \frac{k}{4\pi} A_1 \wedge dA_1 + \frac{k}{2\pi} A_1 \wedge f_2$ from the variation of the total action with respect to $f_2$, compatible with the transformations $\delta A_1 = \lambda_1$, $\delta b_3 = - d \bigl( \rho_2 - \frac{k}{4\pi} \lambda_1 \wedge A_1 \bigr)$.

\section{More examples}
\label{sec:more examples}

We present here two other examples that can be derived by dynamically gauging a $U(1)$ 0-form symmetry.

\subsection{Abelian 2-group symmetry in 4d}

A 0-form and a 1-form symmetry can combine into one algebraic structure known as a 2-group \cite{Kapustin:2013uxa, Gaiotto:2014kfa, Cordova:2018cvg, Benini:2018reh}. Four-dimensional theories with a continuous Abelian 2-group symmetry were discussed in \cite{Cordova:2018cvg}. Consider a 4d Abelian gauge theory coupled to chiral fermions. The photon has a magnetic 1-form symmetry $U(1)^{(1)}$ whose current is $J_2 = * \, (\cF/2\pi)$ written in terms of the dynamical field strength $\cF$. This current is topologically conserved: \mbox{$d * J_2 = d \cF/2\pi = 0$}. The chiral fermions transform under a 0-form symmetry $U(1)^{(0)}$ with current $J_1$. Suppose that there is a gauge-flavor-flavor triangle anomaly, so that $d * J_1 = \frac{k}{4\pi^2} \, (dA_1) \wedge \cF$ where $A_1$ is the background gauge field coupled to $J_1$. It is easy to arrange the charges such that there are no other anomalies, for instance:
\be
\begin{array}{rcccc}
\text{Weyl fermions} : & \psi_1 & \psi_2 & \psi_3 & \psi_4 \\
\text{gauge charges } q_i: & 1 & -1 & 1 & -1 \\
\text{flavor charges } f_i: & 2 & 1 & -2 & -1
\end{array}
\ee
One checks that $\sum q_i^3 = \sum q_i^2 f_i = \sum f_i^3 = 0$, while $\sum q_i f_i^2 \equiv 2k = 6$ in this example
\footnote{\label{f:quantization}The minimal mixed $U(1)$ Chern--Simons terms that are well defined are $\frac1{4\pi^2} A \, dB \, dB$ or $\frac1{8\pi^2} ( A \, dB \, dB + B \, dA \, dA)$.}.
The theory has therefore a modified conservation equation:
\be
d \, {*} J_1 - \frac{k}{2\pi} \, (dA_1) \wedge * J_2 = 0 \,,\qquad d \, {*}J_2 = 0 \,.
\ee
This type of symmetry is called a 2-group symmetry. If we couple this symmetry to $U(1)$ background gauge fields $A_1$ and $C_2$ through the Lagrangian terms
\be
\cL_4 \,\supset\, (*J_1) \wedge A_1 + (*J_2) \wedge C_2 \,,
\ee
the modified conservation equations imply modified gauge transformations for the background fields:
\begin{align}
\delta A_1 &= d\lambda_0 \,,&
\delta b_3 &= d\gamma_2 - \tfrac{k}{2\pi} \, \xi_1 \wedge dA_1 \nn \\
\delta C_2 &= d\eta_1 - \tfrac{k}{2\pi} \, \lambda_0 \, dA_1 \,,\quad&
\delta h_2 &= d\xi_1 \,.
\end{align}
Here we also included the conjugated $\bR$ gauge fields $b_3$ and $h_2$ necessary to write a 5d Lagrangian. A gauge-invariant 5d Symmetry TFT action we can write is:
\be
\label{SymTFT 2-group}
\cZ = \frac{i}{2\pi} \!\int \biggl[ b_3 \wedge dA_1 + h_2 \wedge dC_2 + \frac{k}{2\pi} h_2 \wedge A_1 \wedge dA_1 \biggr] .
\ee
This is the Symmetry TFT for a 2-group symmetry.

Indeed, we can derive this theory from the Symmetry TFT of the free-fermion theory by the procedure of gauging the $U(1)$. The Symmetry TFT of the free-fermion theory is
\be
\cZ = \frac{i}{2\pi} \!\int \biggl[ b_3 \wedge dA_1 + f_3 \wedge dG_1 + \frac{k}{2\pi} G_1 \wedge dA_1 \wedge dA_1 \biggr]
\ee
where $A_1$ and $G_1$ are the $U(1)$ background fields for the flavor symmetry and the to-be gauge symmetry, respectively. Gauging the symmetry is implemented by the replacement $dG_1 \mapsto h_2$ and $f_3 \mapsto dC_2$, as in (\ref{eq:procedure}), which indeed produces the action in (\ref{SymTFT 2-group}).

\subsection{\tps{\matht{\bQ/\bZ}}{Q/Z} non-invertible symmetry in 4d}

Using the building blocks introduced so far, we can derive the Symmetry TFT describing the non-invertible chiral symmetry of QED-like theories \cite{Choi:2022jqy, *Cordova:2022ieu}, which provides an intrinsic definition of an Abelian symmetry with ABJ anomaly. We start from the 5d Symmetry TFT for two 0-form symmetries $U(1)_{\text{A}}\times U(1)_{\text{V}}$ with a mixed AVV anomaly and a pure AAA anomaly:
\begin{align}
\label{eq:symmTFT mixed anomaly}
\cZ &= \frac{i}{2\pi} \!\int_{X_5} \biggl[ b_3 \wedge dA_1 + c_3 \wedge dV_1 \\
&\qquad\quad + \frac{l}{4\pi} \, A_1 \wedge dV_1 \wedge dV_1 + \frac{k}{12\pi} \, A_1 \wedge dA_1 \wedge dA_1 \biggl] . \nn
\end{align}
We assumed that the VAA triangle anomaly vanishes, therefore $l$ must be even. For $l=k=2$ this can be thought of as, for instance, the Symmetry TFT for a Dirac fermion in four dimensions.

The symmetry $U(1)_{\text{V}}$ has no pure anomaly, hence it can be gauged dynamically by coupling it to a photon. This is implemented in the Symmetry TFT as explained in Section~\ref{sec:dynamical gauging}, and the net result is to replace $dV_1 \mapsto f_2$ and $c_3 \mapsto dG_2$. We obtain:
\begin{align}
\label{eq:symmTFT Q/Z}
\cZ &= \frac{i}{2\pi} \!\int_{X_5} \biggl[ b_3 \wedge dA_1 + f_2 \wedge dG_2 \\
&\qquad\quad + \frac{l}{4\pi} \, A_1 \wedge f_2 \wedge f_2 + \frac{k}{12\pi} \, A_1 \wedge dA_1 \wedge dA_1 \biggl]. \nn
\end{align}
We propose that this is the Symmetry TFT for the non-invertible $\bQ/\bZ$ chiral symmetry in 4d.

Let us study this theory more carefully. For simplicity we set $l=2$ (the generalization to other even values of $l$ being straightforward). While the gauge transformations of $b_3$ and $G_2$ are standard, the presence of the non-derivative term $A_1 \wedge f_2\wedge f_2$ forces the gauge transformations of $f_2$ and $A_1$ to also act on $b_3$ and $G_2$:
\begin{align}
\label{eq:modified gauge}
\delta f_2 &= d\lambda_1 \,,\;&
\delta b_3 &= - \tfrac{2}{4\pi} \, \lambda_1 \wedge d\lambda_1 - \tfrac{2}{2\pi} \, \lambda_1 \wedge f_2 \,, \\[.2em]
\delta A_1 &= d\rho_0 \,,\;&
\delta G_2 &= - \tfrac{2}{2\pi} \, \rho_0 \, (f_2 + d\lambda_1) - \tfrac{2}{2\pi} \, \lambda_1\wedge A_1 \,. \nn
\end{align}
Keeping this into account, we analyze the operator content of the theory. First we have 
\be
V_\alpha[\gamma_2] = e^{i\alpha \!\int_{\gamma_2} \! f_2} \,,\qquad W_n[\gamma_1] = e^{in \!\int_{\gamma_1} \! A_1} \,,
\ee
which are both topological and gauge invariant because of the EOMs $dA_1=0$ and $df_2=0$. On the other hand, the integrals of $b_3$ and $G_2$ do not lead to gauge-invariant operators because of \eqref{eq:modified gauge}.

We can try to construct non-genuine operators:
\bea
\label{eq:non-genuine}
\widetilde{U}_\alpha [\gamma_3, D_4] &= \exp \biggl( i\alpha \!\int_{\gamma_3} \! b_3 + i \, \frac{2\alpha}{4\pi} \!\int_{D_4} f_2\wedge f_2 \biggr) , \\
\widetilde{\cT}_n [\gamma_2, D_3] &= \exp \biggl( in \!\int_{\gamma_2} \! G_2 + i \, \frac{2n}{2\pi} \int_{D_3} A_1 \wedge f_2 \biggr) .
\eea
They depend on the open regions $D_4$ and $D_3$, whose boundaries are $\gamma_3$ and $\gamma_2$, respectively. They are gauge invariant and topological because of the EOMs. On a boundary condition which is Dirichlet for $A_1$ and $G_2$, that correspond to the QED-like theory, the boundary values of $A_1$ and $G_2$ play the role of backgrounds fields for the would-be axial $U(1)_\text{A}$ and the magnetic $U(1)^{(1)}$ symmetry, respectively. The endlines of \raisebox{-1.5pt}[0pt][0pt]{$\widetilde{\cT}_n$} would seem to be 't~Hooft lines, charged under the operators $V_\alpha$, while the \novbox{$\widetilde{U}_\alpha$} lying on the boundary would seem to generate the axial symmetry, whose charged operators are the endpoints of $W_n$. However this conclusion is not correct. In the definition \eqref{eq:non-genuine}, $\gamma_3$ and $\gamma_2$ are boundaries and hence homologically trivial. Since the operators are topological, they are essentially trivial and cannot be used to define boundary conditions. Moreover, differently from other cases considered above, the non-genuine operators \novbox{$\widetilde{U}_\alpha$} do not become genuine even on the boundary, since $f_2$ is not set to zero there. One can however do better.

Consider $\widetilde{\cT}_n$ first. The bulk term $\frac{2n}{2\pi}A_1\wedge f_2$ can be thought of as the inflow action for a 2d pure $\bZ_{2n}$ gauge theory, where $A_1$ and $f_2$ are viewed as the background fields for the $\bZ_{2n}$ 0-form and 1-form symmetry, respectively. This implies that if we take $D_3$ to be a tube $\gamma_2 \times [0,1]$, we can place $\exp \bigl( in \!\int_{\gamma_2} \! G_2 \bigr)$ on one end of the tube, and the 2d $\bZ_{2n}$ gauge theory coupled to $A_1$ and $f_2$ on the other end. Then we shrink the tube, so as to define a genuine 2d operator:
\be
\cT_n[\gamma_2] = \widetilde{\cT}_n[\gamma_2] \; \bZ_{2n}[\gamma_2; A_1,f_2] \,.
\ee

We can perform a similar operation with $\widetilde{U}_\alpha$. Since $\alpha$ is a continuous parameter, we should be careful. Indeed a 3d TQFT on which \novbox{$\frac{2\alpha}{4\pi} \!\int_{D_4} \! f_2\wedge f_2$} can topologically terminate only exists if $2\alpha = p/q\in \bQ$ (we take $\gcd(p,q)=1$)
\footnote{For $\alpha = 1/2$ we can simply use the $U(1)_1$ trivial spin-Chern--Simons theory. Indeed this is the only value that corresponds to an invertible symmetry, namely $(-1)^F$.}.
This is the minimal TQFT with 1-form symmetry $\bZ_q$ and anomaly $p$ introduced in \cite{Hsin:2018vcg} and denoted by $\cA^{q,p}$. By coupling its 1-form symmetry to $f_2$ we are able to define a genuine 3-dimensional topological operator
\be
U_{p/2q}[\gamma_3] = \widetilde{U}_{\alpha=p/2q}[\gamma_3] \; \cA^{q,p}[\gamma_3; f_2] \,.
\ee
The operators at irrational $\alpha$, on the other hand, remain non-genuine. 

The procedure we illustrated is the bulk analog of Ref.~\cite{Choi:2022jqy, Cordova:2022ieu}, with the novelty that not only the generators of the chiral symmetry, but also the 2-dimensional operators $\cT_n$ whose endlines are the 't Hooft lines, require  dressing with a TQFT in order to be well defined. This fact is harder to deduce directly from the boundary because the 't~Hooft lines are not topological. One of the advantages of the Symmetry TFT description is that even the topological aspects of non-topological charged objects can be derived from properties of the topological operators in the bulk. Our result is an example of this phenomenon.

The necessity of dressing even the 't~Hooft lines is not surprising, since the non-invertibility of the $\bQ/\bZ$ symmetry is precisely encoded in the action on 't~Hooft lines. When a line crosses the symmetry defect it emerges with a 2-dimensional topological operator attached \cite{Choi:2022jqy, Choi:2022fgx}. It would be interesting to derive this action from the bulk, studying in detail the braiding and crossing of the various operators we introduced.

Finally let us mention that $U_{p/2q}[\gamma_3]$ can be further dressed with an other genuine 3-dimensional operator corresponding to a Chern--Simons term for $A_1$, as we did in Section~\ref{sec:chiral 4d}:
\be
U_{(p/2q,m)}[\gamma_3]=U_{p/2q}[\gamma_3] \times e^{i \frac{m}{4\pi} \!\int_{\gamma_3} A_1\wedge dA_1} \,.
\ee
Studying the triple linking of these operators allows one to characterize the anomaly of the non-invertible chiral symmetry, for which the term $A_1\wedge dA_1\wedge dA_1$ in \eqref{eq:symmTFT Q/Z} is responsible.

\section{Discussion}

Let us mention two open questions. First, is it possible to generalize the construction to non-Abelian 0-form symmetries? A natural candidate for the Symmetry TFT is the non-Abelian BF theory of Ref.~\cite{Horowitz:1989ng}:
\be
\label{non-Abelian action}
\cZ = \frac{i}{2\pi} \int_{X_{d+1}} \! \Tr_\fg \bigl( b_{d-1} \wedge F_2 \bigr) \,.
\ee
Here the fundamental fields are a standard connection $A$ for the Lie group $G$, and a collection $b_{d-1}$ of as many $\bR$ $(d-1)$-form gauge fields as $\dim(\fg)$ ($\fg$ the Lie algebra of $G$) that transform in the adjoint representation of $G$. Then $F_2 = dA - i A\wedge A$ is the non-Abelian field strength of $A$, and $\Tr_\fg$ is the Killing form on $\fg$
\footnote{In fact, one could use a $b_{d-1}$ that takes values in $\fg^*$, hence the trace is not used and $\fg$ does not need to be semisimple.}.
Thus the theory has two sets of gauge transformations:
\be
A \to UAU^{-1} - i \, dU \, U^{-1} \,,\quad b_{d-1} \to U b_{d-1} U^{-1} \,,
\ee
where $U$ takes values in $G$, and
\be
b_{d-1} \to b_{d-1} + D\lambda_{d-2}
\ee
where $D$ is the covariant derivative constructed with $A$, while $\lambda$ is a globally-defined section of the adjoint bundle defined by $A$. The EOMs are $F_2=0$ and $Db_{d-1} = 0$, therefore the theory has topological Wilson line operators
\be
W_\fR[\gamma_1] = \Tr_\fR \, \operatorname{Pexp} \bigl( \mbox{$ i \int_{\gamma_1} A$} \bigr)
\ee
where $\fR$ are representations of $G$. We expect that there exists a topological boundary where these lines can end, and describe local operators transforming in representations $\fR$ of the symmetry $G$. On the other hand, the non-Abelian symmetry defects of the boundary theory are charged under themselves, therefore we do not expect them to exist as genuine topological bulk operators. One possibility is to obtain them from genuine bulk defects labeled by conjugacy classes, that acquire new labels on the boundary \cite{Bonetti:2024cjk}. Another possibility is to realize them as non-genuine bulk defects labeled by group elements, similarly to \cite{Argurio:2024oym}.

It is also natural to expect how to describe anomalies. For instance, for $d$ even, perturbative anomalies will be described by $(d+1)$-dimensional Chern--Simons terms of $A$. For $d=4$, Witten's nonperturbative $SU(2)$ anomaly \cite{Witten:1982fp} will be described by the five-dimensional topological invariant given by the mod-2 index of the Dirac operator. We believe that the theory in (\ref{non-Abelian action}) deserves more study.

Second, the study of TQFTs with an infinite (possibly uncountable) number of objects opens the possibility for new mathematical studies and new physical phenomena
\footnote{See \eg{} Ref.~\cite{Antinucci:2024bcm} for a discussion on the well-definedness of the partition functions of those TQFTs.}.
For instance, as observed in Section \ref{sec:chiral 2d}, even 3d TQFTs with Lagrangian algebras can admit condensations that are not maximal, and cannot be made such. Thus, new TQFTs without gapped boundaries can be derived from theories with gapped boundaries. It would be interesting to explore the consequences of this simple observation for the physical QFTs at the boundary.

\begin{acknowledgments}

We are grateful to Diego Delmastro, Giovanni Galati, Andrea Grigoletto, André Henriques, Pierluigi Niro, Giovanni Rizi, Marco Venuti, and Matthew Yu for useful discussions. We are partially supported by the ERC-COG grant NP-QFT No.~864583 ``Non-perturbative dynamics of quantum fields: from new deconfined phases of matter to quantum black holes", by the MUR-FARE grant EmGrav No.~R20E8NR3HX ``The Emergence of Quantum Gravity from Strong Coupling Dynamics", by the MUR-PRIN grant No.~2022NY2MXY, as well as by the INFN ``Iniziativa Specifica ST\&FI".

\end{acknowledgments}


\setlength{\bibsep}{2pt plus 0.3ex}

\bibliography{TopGravity_PRX}

\end{document}